\documentclass[
superscriptaddress,
twocolumn,
longbibliography,
pra,
aps,
10pt
]{revtex4-1}
\usepackage{graphicx}
\usepackage{amsmath,amsfonts,amssymb}
\usepackage{xcolor}
\usepackage{hyperref}
\hypersetup{colorlinks=true,allcolors=blue}
\usepackage[T1]{fontenc}
\usepackage{footnote}
\usepackage{siunitx}
\usepackage{physics}
\usepackage{dsfont}
\usepackage{tabularx,array}

\usepackage{xcolor}
\definecolor{color_g}{HTML}{2577B0}
\definecolor{color_dm}{HTML}{277A28}
\definecolor{color_bm}{HTML}{FD822B}
\definecolor{color_b}{HTML}{B4B4B4}
\definecolor{color_bp}{HTML}{B4B4B4}
\definecolor{color_dp}{HTML}{B4B4B4}
\definecolor{color_text}{HTML}{B4B4B4}

\renewcommand{\t}[1]{\textrm{#1}}
\newcommand{\nn}{\nonumber\\}

\newcommand{\+}{^\dagger}
\renewcommand{\>}{\rangle}
\newcommand{\<}{\langle}

\newcommand{\g}{\gamma}

\renewcommand{\P}{\mathcal{P}}
\newcommand{\I}{\mathcal{I}}

\renewcommand{\O}{\mathcal{O}}

\begin{document}

%\title{Off-resonant excitation of quantum systems for single-photon generation}
\title{Swing-up of quantum emitter population using detuned pulses}
% pulses tuned below the resonance ? 
%SUPER = Swing UP of quantum emittER
% Affiliations
%\newcommand{\WWU}{Institute of Solid State Theory, University of Münster, 48149 Münster, Germany}

\author{Thomas K. Bracht}
%\affiliation{Institute of Solid State Theory, University of Münster, 48149 Münster, Germany}
\affiliation{Institut f{\"u}r Festk{\"o}rpertheorie, Universit{\"a}t M{\"u}nster, 48149 M{\"u}nster, Germany}
\author{Michael Cosacchi}
\author{Tim Seidelmann}
\affiliation{Theoretische Physik III, Universit{\"a}t Bayreuth, 95440 Bayreuth, Germany}
\author{Moritz Cygorek}
\affiliation{Heriot-Watt University, Edinburgh EH14 4AS, United Kingdom}
\author{Alexei Vagov}
\affiliation{Theoretische Physik III, Universit{\"a}t Bayreuth, 95440 Bayreuth, Germany}
\affiliation{ITMO University, St. Petersburg, 197101, Russia}
\author{V. Martin Axt}
\affiliation{Theoretische Physik III, Universit{\"a}t Bayreuth, 95440 Bayreuth, Germany}
\author{Tobias Heindel}
\affiliation{Institut f{\"u}r Festk{\"o}rperphysik, Technische Universit{\"a}t Berlin, 10623 Berlin, Germany}
\author{Doris E. Reiter}
\affiliation{Institut f{\"u}r Festk{\"o}rpertheorie, Universit{\"a}t M{\"u}nster, 48149 M{\"u}nster, Germany}

\date{\today}

\begin{abstract}
The controlled preparation of the excited state in a quantum emitter is a prerequisite for its usage as single-photon sources - a key building block for quantum technologies. In this paper we propose a coherent excitation scheme using off-resonant pulses. In the usual Rabi scheme, these pulses would not lead to a significant occupation. This is overcome by using a frequency modulated pulse to swing up the excited state population. The same effect can be obtained using two pulses with different strong detunings of the same sign. We theoretically analyze the applicability of the scheme to a semiconductor quantum dot. In this case the excitation is several meV below the band gap, i.e., far away from the detection frequency allowing for easy spectral filtering, and does not rely on any auxiliary particles such as phonons. Our scheme has the potential to lead to the generation of close-to-ideal photons.
\end{abstract}

\maketitle

\section{Introduction}
Deterministically preparing the excited state of a quantum emitter is a key to many applications in quantum information technology, since the subsequent decay of the excited state yields a single-photon \cite{senellart2017high,aharonovich2016,rodt2020deterministically}. Prominent examples for quantum emitters are semiconductor quantum dots \cite{Michler2000,Santori2001,He2013,Somaschi2016,wei2014deterministic,Ding2016,zhai2021quantum}, strain potentials and defects in monolayers of atomically thin semiconductors \cite{tonndorf15single,chakraborty2019advances,klein2019side}, defect centers in diamond \cite{Englund2010,Beha2012,Fehler2019,
Janitz2020,Schrinner2020} or in hexagonal boron nitride \cite{Proscia2020,Froech2020,grosso2017tunable}. The deterministic preparation relies on the direct excitation of the quantum emitter excited state by an external laser pulse. Since the photon emission is controlled by the timing of the laser pulse, the source is called on-demand or deterministic. A common way to achieve a precise preparation is using resonant excitation yielding Rabi rotations \cite{stievater2001rabi,kamada2001exciton,He2013,he2017deterministic,reiter2014role}. However, this scheme suffers from several drawbacks: Because the excitation and detection is at the same frequency, sophisticated filtering has to be applied to separate the desired photons. Additionally, this scheme is sensitive to variations in the pulse parameters. The latter can be overcome by using chirped pulses within the adiabatic rapid passage (ARP) scheme, though the necessity to apply filtering remains \cite{simon2011robust,wei2014deterministic,kaldewey2017coherent}. Another direct excitation scheme is the phonon-assisted preparation, where the laser is tuned above the exciton resonance and the exciton state is then occupied by phonon-induced relaxation \cite{Ardelt2014,bounouar2015phononassisted,quilter2015phononassisted,barth2016fast,cosacchi2019emission}. This overcomes the problem of exciting and measuring at the same frequency, but the drawback of this scheme is that it relies on an incoherent relaxation path. Excitations with a strongly detuned laser pulse beyond the phonon spectral density will only lead to a vanishing transient occupation of the quantum emitter.

In this article an alternative scheme is proposed that results in the Swing-UP of the quantum EmitteR population (SUPER). The SUPER scheme makes use of highly detuned laser pulses, which in a classical Rabi scheme would not lead to any significant excited state occupation at all.
Our scheme relies on periodic changes of the Rabi frequency, leading to a swing-up behavior in the occupation dynamics. We explain the mechanism of the SUPER scheme using a frequency modulation (FM) of the exciting pulse. An implementation of the SUPER scheme with state-of-the-art lasers can be achieved using a two-color protocol. We show that this two-color protocol leads to a complete exciton occupation and near-unity single-photon purity, indistinguishability and photon output. This makes the SUPER scheme ideal for implementing a single-photon source. 

\section{Concept of the SUPER scheme}
\begin{figure}
    \includegraphics{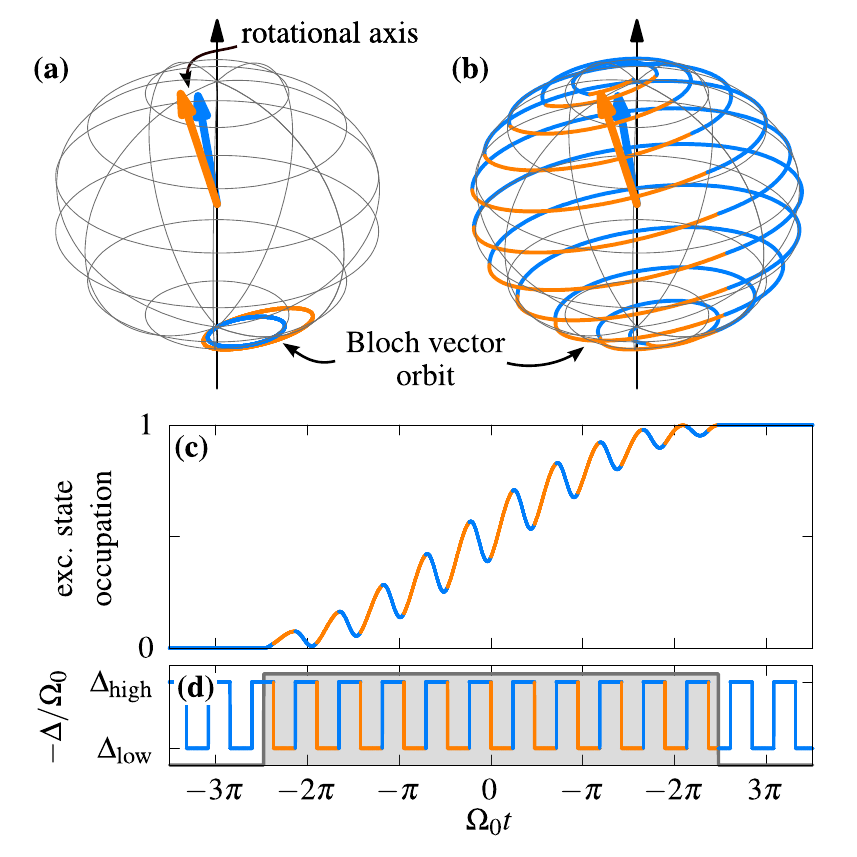}
    \caption{Bloch vector dynamics for (a) off-resonant Rabi oscillations (b) the SUPER scheme. (c) Dynamics of the excited state occupation for the SUPER scheme under (d) a time-dependent detuning switching between $\Delta_{\text{high}}=\SI{-5.47}{\Omega_0}$ and $\Delta_{\text{low}}=\SI{-2.74}{\Omega_0}$. The pulse has a duration of $4.95\pi/\Omega_0$.}
    \label{fig:dynamics_rectangular}
\end{figure}

To grasp the concept of the swing-up scheme, it is most instructive to look at a pulse which is modulated in a step-like manner. We consider a generic two-level system with the energy spacing $\hbar\omega_0$. For constant driving with laser frequency $\omega_L$, the two-level system performs Rabi oscillations with frequency $\Omega_R$ and amplitude $a$
\begin{equation}
    \Omega_R = \sqrt{\Omega_0^2 + \Delta^2}, \quad a= (\Omega_0/\Omega_R)^2,
    \label{eq:rabifreq}
\end{equation}
where $\Omega_0$ is proportional to the laser amplitude and 
\begin{equation}
\Delta=\omega_L-\omega_0
\end{equation}
is the spectral detuning. We note that only a resonant pulse, i.e., $\Delta=0$, with pulse area $\alpha=(2n+1)\pi$, $n\in\mathbb{N}_0$, results in complete inversion of the system. When the detuning is large, the amplitude of the Rabi oscillations becomes negligible. They can be displayed on the Bloch sphere as shown in Fig.~\ref{fig:dynamics_rectangular}(a). 
The Bloch vector makes a circular motion around its rotational axis (arrows). Due to the large detuning, it stays near the south pole of the sphere, corresponding to low excitation. The two examples show the detunings $\Delta_{\text{low}}=\SI{-2.74}{\Omega_0}$ (orange) and $\Delta_{\text{high}}=\SI{-5.47}{\Omega_0}$ (blue). 

Interestingly, if we combine these two off-resonant oscillations in a clever way, we can reach a full inversion of the quantum emitter. This can be seen in Fig.~\ref{fig:dynamics_rectangular}(b,c), which show Bloch vector and temporal dynamics of the SUPER scheme using a rectangular pulse.\\
We achieve the population inversion by switching back and forth between the two detunings $\Delta_{\text{high}}$ and $\Delta_{\text{low}}$. 
Each time the occupation of the excited state increases, we use the smaller detuning $\Delta_{\text{low}}$ (higher amplitude of the Rabi oscillation). On the other hand, if the occupation falls, we use the higher detuning $\Delta_{\text{high}}$ (lower amplitude of the Rabi oscillation). By this precise timing,  each oscillation period in (d) gives rise to a small increase in population of the excited state and the occupation swings up to the excited state.

The frequency of the modulation in the SUPER scheme typically lies close to the Rabi frequency $\Omega_R^{(\text{C})}=\sqrt{\Omega_0^2+\Delta_{\text{C}}^2}$ induced by a constant pulse with the mean detuning $\Delta_{\text{C}}= (\Delta_{\text{high}}+\Delta_{\text{low}})/2$. Looking more closely, we find that an occupation of unity can reliably be achieved by using a slightly longer time at $\Delta_{\text{low}}$ than $\Delta_{\text{high}}$, according to the Rabi frequencies corresponding to these detunings. 

The spectrum of the driving pulse contains several peaks, one of those close to the transition frequency $\omega_0$. One could, therefore, expect that the driving is in resonance with the transition and this is the reason for the effective occupation change. This is, however, not the case. Using Gaussian pulses for the excitation we demonstrate below that SUPER works, even when none of the spectral peaks of the driving pulse are in resonance with the two-level system transition. 

%This means, that the modulation switches the detuning with the Rabi frequency, using the mean detuning $\Delta=\SI{-4.10}{\Omega_0}$ in Eq.~\ref{eq:rabifreq} due to simplicity. 

%%%%%%%%%%%%%%%%%%%%%%%%%%%%%%%%%%%%%%%%%% FM SUPER
\section{FM-SUPER scheme}\label{sec:fm_case}
We now consider the excitation of a specific system, namely a quantum dot excited by an optical laser pulse. Semiconductor quantum dots have already been shown to perform well as deterministic single-photon sources \cite{Ding2016,Somaschi2016,senellart2017high,Wang2020,Arakawa2020,
Tomm2021,Thomas2021b, zhai2021quantum}, for which a high-fidelity preparation of the excited state is necessary. In such a system, the energy separation of ground and excited state is in the range of $1-2\,\si{\electronvolt}$ and typical detunings of the laser are of the order of several meV. Therefore, the temporal dynamics takes place on a picosecond time scale. We note that for phonon-assisted schemes detunings of $1-4\,\si{meV}$ are typical \cite{bounouar2015phononassisted,quilter2015phononassisted,Ardelt2014}, which is already sufficient to perform filtering of the exciting laser for the detection process. In our scheme we consider highly detuned pulses with detunings larger than $\SI{5}{meV}$. Note that smaller detunings might be possible using longer pulses. 

In the optical regime, experimental realizations often rely on laser pulses with a Gaussian pulse shape.
Therefore we consider pulses of the form
\begin{equation}
  \Omega(t) = \Omega_0(t) e^{-i\phi(t)}, \quad\Omega_0(t) = \frac{\alpha}{\sqrt{2\pi\sigma^2}} e^{-t^2/(2\sigma^2)}\label{eq:electric_field}
\end{equation}
 where $\Omega_0(t)$ is the pulse envelope with pulse area $\alpha=\int\Omega_0(t)\,\text{d}t$ and duration $\sigma$.
 The time-dependent detuning is connected to the phase of the laser via $\Delta(t) = \dot{\phi}(t)-\omega_0$.

Motivated by the periodic switching of the detuning leading to the swing-up effect for the rectangular pulse, we use FM assuming a smooth switching process resulting in the general form of a sinusoidal detuning
\begin{equation}
    \Delta(t) = \Delta_{\text{C}} + \Delta_{\text{M}} \sin(\omega_{\text{M}}\,t) \label{eq:modulation_freq}\,.
\end{equation}
The laser frequency is composed of a constant detuning $\Delta_{\text{C}}$ and a sinusoidal modulation with amplitude $\Delta_{\text{M}}$ and frequency $\omega_{\text{M}}$. Similar to the considerations using the rectangular pulse, we expect a good performance of the FM-SUPER scheme if the modulation frequency $\omega_M$ is close to the Rabi frequency at the pulse maximum, i.e. $\omega_M \approx \sqrt{\Omega_0^2 + \Delta_\text{C}^2}$. 

\begin{figure*}
    \centering
    \includegraphics{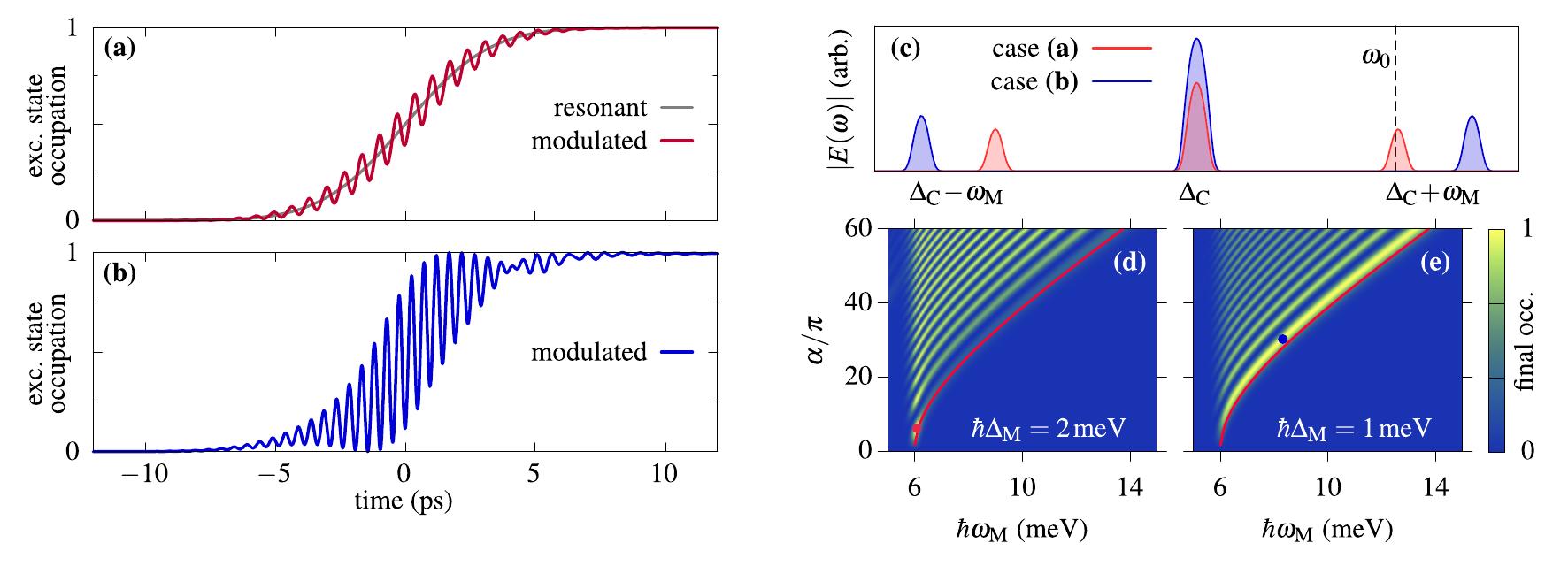}
    \caption{Time evolution of the excited state occupation for (a) $\hbar\Delta_{\text{M}}=\SI{2}{meV}, \alpha=6.2\pi$ and (b) $\hbar\Delta_{\text{M}}=\SI{1}{meV}, \alpha=30.3\pi$ \\{}[red/blue dot in (d/e)]. (c) Spectra of the pulses used in (a,b), the dashed line marks the excited state energy. (d,e) Final excited state occupation for different modulation parameters and pulse areas with $\hbar\Delta_{\text{C}}=\SI{-6}{meV}$ and $\sigma=\SI{4}{ps}$. The red lines mark the Rabi frequency at the pulse maximum.}
    \label{fig:fm_multiplot}
\end{figure*}
Figure~\ref{fig:fm_multiplot}(a,b) exemplarily shows the dynamics resulting from two different parameter sets. In both cases, we use a constant detuning of $\hbar\Delta_{\text{C}} = \SI{-6}{meV}$ and $\sigma=4\,\text{ps}$, with $\hbar\Delta_{\text{M}} = \SI{2}{meV},\alpha=6.2\pi, \hbar\omega_{\text{M}}=\SI{6.08}{meV}$ for (a) and $\hbar\Delta_{\text{M}} = \SI{1}{meV},\alpha=30.3\pi, \hbar\omega_{\text{M}}=\SI{8.32}{meV}$ for (b). In both cases the occupation is completely transferred to the excited states, while performing small oscillations. This demonstrates that the SUPER scheme also works for Gaussian pulses, even for high detunings where without modulation no occupation of the excited state would occur.

Due to the frequency modulation, the spectrum of the pulse does not only contain the carrier frequency leading to the detuning $\Delta_{\text{C}}$, but also additional side-bands at multiples of the modulation frequency $\omega_{\text{M}}$. The moderate pulse duration of $\sigma=\SI{4}{ps}$ leads to a small contribution to the spectral width by the pulse envelope, so the distinct peaks in the spectrum are relatively sharp. This is shown in the spectra in Fig.~\ref{fig:fm_multiplot}(c), corresponding to the cases shown in (a,b). The question now arises, which role these side-bands play in the SUPER scheme. 

Of most interest is the first side-band, which can be resonant to the ground-state exciton transition if $|\omega_{\text{M}}|\approx\Delta_{\text{C}}$. To get insight into the relevance of $\omega_{\text{M}}$, we analyze the final excited state occupation, i.e., the occupation after the pulse for different parameters. Accordingly, Fig.~\ref{fig:fm_multiplot}(d,e) shows the impact of $\omega_{\text{M}}$ for different pulse areas $\alpha$, for (d) $\hbar\Delta_{\text{M}}=\SI{2}{meV}$ and (e) $\hbar\Delta_{\text{M}}=\SI{1}{meV}$. In both cases a stripe pattern sets in at $\hbar\omega_{\text{M}}=\SI{6}{meV}$. We remind the reader that $\hbar\Delta_{\text{C}}=\SI{-6}{meV}$ was used, so this parameter set leads to a resonant side-band. This is mainly responsible for the final occupation in this regime. Indeed, approximating this first side-band amplitude by the Bessel function of the first kind leads to an effective pulse area of $\alpha_{\text{eff}} = \alpha J_{1}\left(\Delta_{\text{M}}/\omega_{\text{M}}\right)$. This breaks the scheme for $\hbar\omega_{\text{M}}\approx\SI{6}{meV}$ down to Rabi rotations using the effective pulse area of the side-band, also resulting in the stripe pattern in $\alpha$. We note that the parameters used in Fig.~\ref{fig:fm_multiplot}(a) mark such a case [red dot in (d)]. Therefore, in Fig.~\ref{fig:fm_multiplot}(a) we also show a resonant excitation, which agrees with the mean of the oscillations.

However, if we increase the pulse area and accordingly $\omega_{\text{M}}>|\Delta_{\text{C}}|$, the first side-band is no longer resonant to the transition frequency $\omega_0$. In these cases, the full occupation of the excited state can no longer be explained by resonant Rabi oscillations any more. Instead, we here truly make use of the SUPER mechanism. 
When moving to higher detunings, the stripes in Fig.~\ref{fig:fm_multiplot}(d,e) show a square-root-like behavior. The lowest stripe is similar to the behavior of the Rabi frequency as a function of pulse area, shown as a red line. Using $\hbar\Delta_{\text{M}}=\SI{1}{meV}$ in (e), this first stripe shows a high final occupation, even for large $\omega_{\text{M}}$. One example using this truly off-resonant SUPER scheme is marked by the blue dot in (e). The modulation with $\hbar\omega_M=\SI{8.32}{meV}$ leads to a side-band located at $\SI{2.32}{meV}$, which is well out of resonance. The time evolution for this case is shown in Fig.~\ref{fig:fm_multiplot}(b) and exhibits high-frequency high-amplitude oscillations in the occupation, ending up in the excited state.

By using the SUPER scheme, we can achieve full occupation of the excited state by a highly detuned pulse, which has no spectral component resonant to the transition frequency. It should be noted that absence of the resonance with the bare transition frequency $\omega_0$ does not exclude resonances with full non-linear dynamics of the system.

Assuming the case of an optically excited quantum dot, the sinusoidal modulation of the laser is on the femtosecond time scale. To the best of our knowledge, such a fast modulation is not possible with state-of-the-art laser technology. We propose a two-color solution in the following, which provides a practical route to exploit the SUPER scheme with standard laser technology. Nonetheless, the FM-SUPER scheme conveys convincingly that a periodic change of the laser frequency can result in a coherent preparation of the excited state and it would be interesting to see if this FM scheme can be realized in other two-level systems like superconducting circuits \cite{Hofheinz2008,Hofheinz2009} .

%%%%%%%%%%%%%%%%%%%%%%%%%%%%%%%%% 2C SUPER
\section{Two color SUPER scheme}\label{sec:am_case}
The SUPER scheme relies on using Rabi oscillations with different frequencies and amplitudes. For constant frequency, an amplitude modulation can likewise result in a change of the Rabi frequency, see Eq.~\eqref{eq:rabifreq}. Such an amplitude modulation can be achieved by the beats induced by two laser pulses. We will consider two Gaussian pulses of similar width but different detunings. Such pulses are straight-forward to realize in an experimental setting, when compared to frequency modulated pulses. 
The electric field consists of two pulses with constant energy to excite the system,
\begin{equation}
    \Omega(t) = \Omega_{1}(t)e^{-i\omega_1 t} + \Omega_{2}(t-\tau)e^{-i\omega_2t + i\varphi},\label{eq:two_color_laser}
\end{equation}
where $\Omega_{j}$ $(j=1,2)$ are the real envelopes of two Gaussian-shaped pulses with (in general) different pulse area $\alpha_{j}$ and duration $\sigma_{j}$ as well as temporal separation of $\tau$.
The frequencies of the lasers will again be expressed in terms of the detuning $\Delta_j = \omega_j - \omega_0$ for each laser pulse. Accordingly, we call this the two-color (2C)-SUPER scheme. Additionally, an arbitrary phase difference $\varphi$ between the pulses is introduced. As shown in appendix \ref{app:phase}, the phase does not alter the final occupation, such that we set $\varphi=0$ in the following.

To achieve a high occupation of the excited state, the swing-up process has to be coordinated, which is again achieved by choosing the two detunings of the laser in such a way that the difference between the detunings is equal to the Rabi frequency associated with the first pulse. Therefore, if we set the detuning for one laser pulse to $\Delta_1$ and use the corresponding Rabi frequency $\Omega_1$ at its maximum, the detuning of the other pulse calculates to
\begin{equation}
    \Delta_2 = \Delta_1 - \sqrt{\Omega_{1}^2(t=0) + \Delta_1^2}. \label{eq:second_frequency}
\end{equation}
From the equation we see that $|\Delta_2| > 2|\Delta_1|$, such that the second pulse is detuned even further from the transition. Note that we use $\Delta_1<0$ in order to obtain an excitation in the transparent region below the exciton line.

\begin{figure}
    \includegraphics{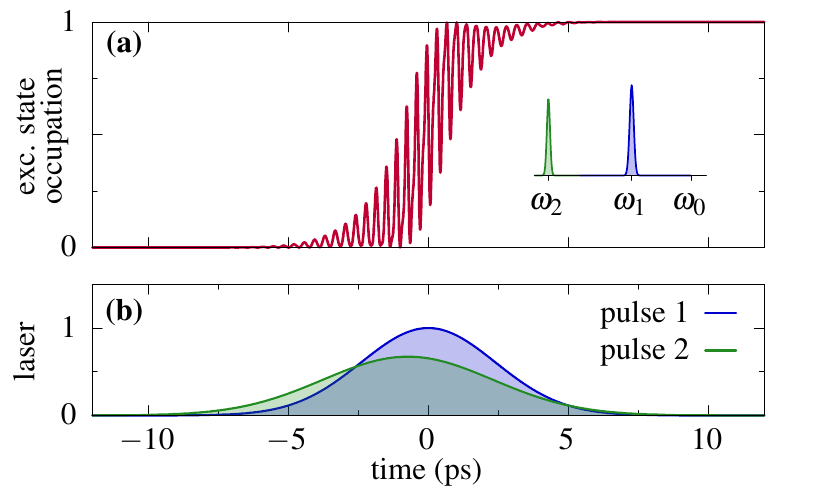}
    % tau1=2400, tau2=3040, area1=22.31*np.pi, area2=19.9*np.pi, t02=-730, detuning=-8.0
    % times in fs, energies in meV
    \caption{(a) Time evolution of the excited state occupation for the 2C-SUPER scheme. The inset shows the combined spectrum of the pulses. (b) Normalized laser envelope for the pulse sequence used. Parameters are given in Tab.~\ref{tab:table_am_parameters} for $\hbar\Delta_1=\SI{-8}{meV}$.
    }
    \label{fig:twocolor_example}
\end{figure}

An example of the occupation dynamics for the 2C-SUPER scheme is shown in Fig.~\ref{fig:twocolor_example}(a). The two pulses, which are both energetically well below the band gap, lead to a complete transfer of the electronic system to the excited state. During the pulses, the occupation performs fast oscillations, similar to the case of FM-SUPER. The right inset shows the spectrum of the complete laser field, each pulse leads to a distinct peak centered around the corresponding frequency, i.e., there is no spectral overlap with the transition energy of the quantum emitter. Here, $\hbar\Delta_1=\SI{-8}{meV}$ and $\hbar\Delta_2=\SI{-19.1630}{meV}$ were used. In Fig.~\ref{fig:twocolor_example}(b) the envelopes of the pulses are shown, having different pulse durations ($\sigma_{1}=\SI{2.4}{ps},\sigma_2=\SI{3.04}{ps}$, corresponding in the spectral regime to a FWHM of \SI{0.65}{meV} and \SI{0.5}{meV} respectively) and pulse areas ($\alpha_1=\SI{22.65}{\pi}, \alpha_2=\SI{19.29}{\pi}$). Additionally both pulses are separated by $\tau=\SI{-0.73}{ps}<0$, which leads to the pulse we refer to as the second pulse arriving earlier in time than the first one, while both pulses still show a strong overlap. 

\begin{figure}
    \centering
    \includegraphics{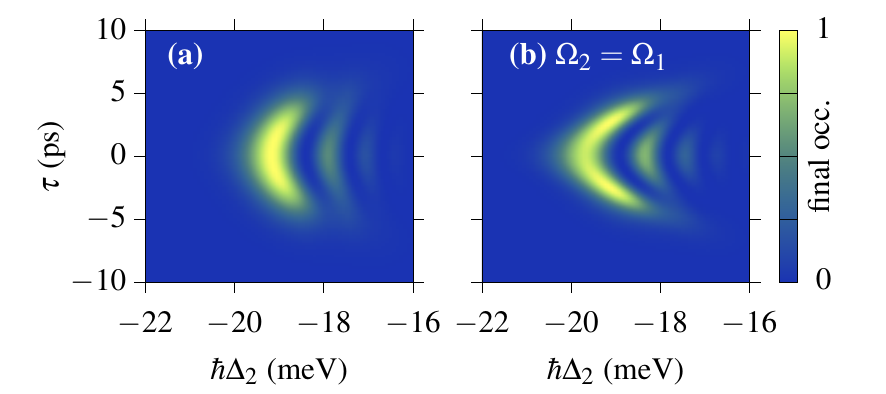}
    % tau1=2400, tau2=3040, area1=22.31*np.pi, area2=19.9*np.pi, detuning=-8.0
    % times in fs, energies in meV
    \caption{Final excited state occupation for different detunings $\Delta_2$ and time delays $\tau$ of the second pulse. For a negative time difference, the second pulse arrives earlier than the first. (a) parameters like in Fig.~\ref{fig:twocolor_example}, (b) same as (a), but the second pulse has the same shape as the first one. 
    }
    \label{fig:twocolor_detune2_timediff}
\end{figure}

Next, we evaluate for which parameters a high fidelity inversion is possible. As such, we start from the parameters used in the previous figure and vary $\Delta_2$ and $\tau$, which is shown in Fig.~\ref{fig:twocolor_detune2_timediff}(a). 
The region where high final occupations are reached is symmetric regarding the sign of the pulse offset $\tau$, such that it does not matter if the second pulse arrives earlier or later than the first pulse. For growing $|\tau|$ the required detuning decreases, i.e., the energy of the second pulse increases slightly. If the offset is too large, the pulses arrive separately from each other and no inversion is achieved for high $|\tau|$.

The pulses used so far had different pulse durations and areas. In Fig.~\ref{fig:twocolor_detune2_timediff}(b) we analyze the final occupation for two pulses of identical shape, where we choose $\sigma_2 = \sigma_1$ and $\alpha_2=\alpha_1$. The result is a behavior similar to that in (a), with a few distinctions: For completely overlapping pulses ($\tau=0$) complete inversion can not be achieved, but the occupation only goes up to a maximum of about \SI{90}{\percent}. Only for a finite pulse separation, e.g., for $\tau\gtrsim\SI{2.5}{ps}$, the 2C-SUPER scheme results in a full occupation of the excited state. 

For the variations of the detuning, we find that in both Fig.~\ref{fig:twocolor_detune2_timediff}(a) and Fig.~\ref{fig:twocolor_detune2_timediff}(b), the detuning $\Delta_2$ which leads to the best results corresponds to that given by Eq.~\eqref{eq:second_frequency}, confirming that the concept of the SUPER scheme depends on the Rabi frequency. Additionally, in both panels (a) and (b) small maxima are visible for larger $\Delta_2$. These stem from excitation and de-excitation processes similar to higher order Rabi rotations. However, here the scheme does not achieve full inversion. 

\begin{figure}
    \centering
    \includegraphics{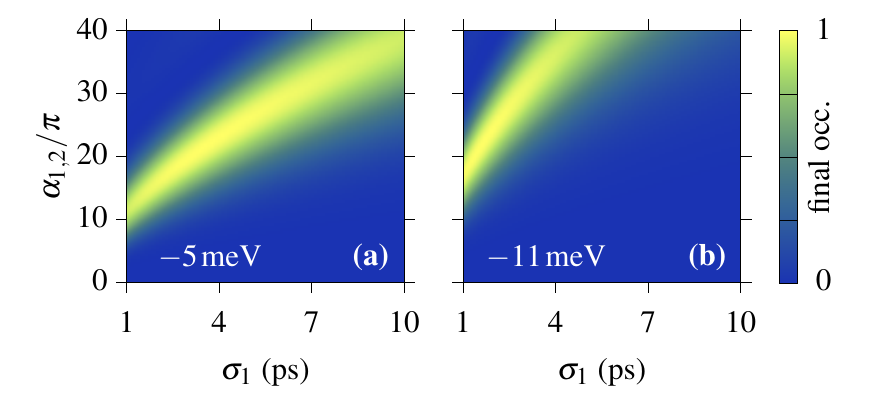}
    \caption{Final excited state occupation for different pulse areas and pulse durations. Here both pulses always have the same pulse area, $\tau = 0$ and $\sigma_2=1.5\sigma_1$. (a) $\hbar\Delta_1=\SI{-5}{meV}$, (b) $\hbar\Delta_1=\SI{-11}{meV}$.
    }
    \label{fig:multiparameter}
\end{figure}

Next, we analyze the impact of the pulse shape on the 2C-SUPER scheme. As such, we vary the pulse areas and the pulse length systematically in Fig.~\ref{fig:multiparameter}. The figure shows the excited state occupation for the case that the pulse areas of both pulses are equal ($\alpha_1=\alpha_2$), but have different length $\sigma_2=1.5\sigma_1$ and $\tau=0$.
In (a) the detuning of the first pulse is set to $\SI{-5}{meV}$, in (b) $\SI{-11}{meV}$ is used. In both cases the scheme yields high inversion over a broad parameter range and the structure of the behavior is qualitatively the same. It is important to note, that the changes in pulse area and pulse width lead to a change of the second detuning, according to Eq.~(\ref{eq:second_frequency}). Looking at the pulse parameters in (b), where larger detunings are used, also larger pulse areas or shorter pulses are necessary for the scheme to work efficiently, which correspond to a larger pulse intensity. The behavior at very small detunings ($\sim \SI{-1}{meV}$) for otherwise the same excitation parameters is discussed in appendix~\ref{app:small_det}. %The dependence between $\alpha$ and $\sigma$ is approximately linear, from which we conclude that for efficient coupling, the laser has to have a certain intensity which is proportional to $\alpha/\sigma$. For the higher detuning in (b) the coupling is weaker so higher pulse areas are needed for the SUPER scheme to work. 

\begin{table*}[t]
{
\caption{Example parameter sets for which the two-color scheme yields a close to unity excited state occupation, using different detunings $\Delta_1$. 
Here, $\Delta_2$ is always calculated using the analytic expression in Eq.~\eqref{eq:second_frequency}.
}
\label{tab:table_am_parameters}
\renewcommand{\arraystretch}{1.5}
\begin{tabularx}{\textwidth}{XXXXXXX}
    \toprule
    $\hbar\Delta_1$ in meV & $\hbar\Delta_2$ in meV & $\sigma_1$ in ps & $\sigma_2$ in ps & $\alpha_1$ & $\alpha_2$ & $\tau$ in ps  \\
    \hline
    $-5.0$ & $-11.7826$ & $4.50$ & $7.10$ & $25.00\pi$ & $25.00\pi$ & $\phantom{-}0.00$  \\
    $-5.0$ & $-13.5624$ & $1.55$ & $1.55$ & $13.06\pi$ & $10.45\pi$ & $\phantom{-}0.00$  \\
    $-5.0$ & $-12.5522$ & $2.23$ & $5.00$ & $15.30\pi$ & $28.31\pi$ & $\phantom{-}2.40$ \\  
    $-8.0$ & $-19.1630$ & $2.40$ & $3.04$ & $22.65\pi$ & $19.29\pi$ & $-0.73$ \\
    %$-8.0$ & $-19.0818$ & $2.40$ & $3.04$ & $22.31\pi$ & $19.90\pi$ & $-0.73$ \\
    $-11.0$ & $-25.7450$& $2.50$ & $3.60$ & $28.57\pi$ & $27.35\pi$ & $\phantom{-}1.50$  \\
    \hline
    \hline
\end{tabularx}
\renewcommand{\arraystretch}{1.0}
}
\end{table*}

This two-color approach to the SUPER scheme performs well on a broad range of parameters. In an experimental environment only two separate laser pulses are required, while there are many possible parameters for which the scheme works. To further illustrate this, we give a few examples of laser parameters which result in an excited state occupation of one in Tab.~\ref{tab:table_am_parameters}. The parameters found are generally accessible with standard laser technology \cite{huber2015optimal}. Likewise, two pulses could be generated by carving from a single femtosecond pulse \cite{koong2021coherent}. Because the final occupation in the scheme does not depend on the phase difference between the pulses (cf. appendix~\ref{app:phase}, Fig.~\ref{fig:twocolor_phase}), phase locking should not be required. 

We find that the required pulse areas used in the SUPER scheme are rather high compared to resonant excitation, leading to high peak intensities of the laser pulses. While high laser intensities can damage the quantum dot, one has to keep in mind that we use pulses detuned below the transition energy. When choosing such a large negative detuning, we are in the transparent region of the material, which results in low absorption-induced heating of the sample. Note that this is different for a positive detuning, where higher excited states or states of the host material could be excited by these pulses. 

%%%%%%%%%%%%%%%%%%%%%%%%%%%%%%%%% Single-photon source
\section{Performance as single-photon source}\label{sec:single_photons}
In terms of quantum technology, one of the most prominent applications of optical state control of two-level systems is their usage as single-photon sources. In the case of quantum dots, single-photon emission has already been achieved in several experiments \cite{Santori2002,Ding2016,Somaschi2016,senellart2017high,Schweickert2018,Wang2020,Arakawa2020,Tomm2021,Thomas2021b, zhai2021quantum}, but the search for the optimal source is ongoing \cite{thomas2021race} including the development of new excitation protocols like ARP \cite{simon2011robust,wei2014deterministic,kaldewey2017coherent} or phonon-assisted schemes \cite{Ardelt2014,bounouar2015phononassisted,quilter2015phononassisted,barth2016fast,cosacchi2019emission} or symmetrically detuned excitation \cite{he2019coherently,koong2021coherent}. To compete with these preparation protocols in terms of photon sources, the SUPER scheme should at least theoretically have a similar performance regarding the photon quality. 

The quality of the generated photons can be estimated in terms of three quantities: 
the single-photon purity $\P$, the indistinguishability $\I$ of two subsequently emitted photons, and the brightness of the source. Especially the brightness can have different definitions, amounting to the number of photons generated \cite{Manson2016,Gustin2018} or collected \cite{Somaschi2016} per excitation. Here, to estimate the brightness, we calculate the photon output $\O$ as the probability of a photon being generated per excitation cycle. A definition of these quantities as well as methods to calculate them are given in appendix~\ref{app:def_photonic}. The source is a perfect single-photon emitter, if all three quantities are unity. We evaluate them for the 2C-SUPER scheme, which is more straightforward to implement experimentally, using the parameters from Fig.~\ref{fig:twocolor_example}. For the calculations we are additionally assuming a radiative decay rate of \SI{1}{\per\nano\second} for the quantum dot exciton. 

The results are displayed in Tab.~\ref{tab:PIB}, where we find that all quantities are very close to unity for the 2C-SUPER scheme. They are compared with the corresponding characteristics of a resonant $\pi$-pulse excitation with otherwise the same pulse parameters. Using the 2C-SUPER scheme, we can reach a single-photon purity and indistinguishability of \SI{99.85}{\percent} as well as a photon output of \SI{99.64}{\percent}, such that our protocol can compete with the single-photon source operated in resonance fluorescence and compares favorably concerning $\P$ and $\I$.
The crucial difference is the off-resonant operation of the quantum dot in our protocol presented in this work, which allows for a spectral separation of pump and signal.
Note that $\P$ and $\I$ actually do differ slightly (beyond the decimal places shown in Tab.~\ref{tab:PIB}). The reason for their similarity lies in the fact that the single-photon purity is already close to unity.

%neue Werte:
%.99848513        .99848559542248951778   .99638184807073615957   .99848606

\begin{table}[t]
\renewcommand{\arraystretch}{1.5}
\begin{center}
\caption{Single-photon characteristics for the 2C-SUPER scheme and a resonant excitation with a Gaussian $\pi$-pulse. For definitions, see appendix~\ref{app:def_photonic}.}
  \begin{tabularx}{\columnwidth}{lXX}
 
    \hline
    \hline	
    & 2C-SUPER & $\pi$-pulse \\
    \hline
    Single-photon purity $\P$ & $99.85\,\%$ & $99.75\,\%$ \\
    Indistinguishability $\I$ & $99.85\,\%$ & $99.75\,\%$ \\
    Photon output $\O$ & $99.64\,\%$ & $99.85\,\%$ \\
    \hline
    \hline
  \end{tabularx}
  \label{tab:PIB}
\end{center}
\renewcommand{\arraystretch}{1.0}
\end{table}

The state preparation and the photon properties are also subject to decoherence mechanisms for example via phonons, the hyperfine interaction or the interaction with nuclear spins. For an optically excited quantum dot, phonons have been shown to be a major source of decoherence, which already might hinder the state preparation schemes drastically \cite{ramsay2010damping,luker2019review,Reiter2019}.
%Focusing on the phonon effect, for values of both $\P$ and $\I$ farther away from unity than in the case studied above, the phonon environment would affect $\I$ more than $\P$ \cite{Cosacchi2021}. 
Therefore, we briefly analyze, if the phonons hinder the preparation of the excited state in the quantum dot when employing the scheme presented here. For this, we use the standard pure dephasing-type Hamiltonian and apply a path-integral formalism to solve the corresponding equations of motions \cite{Vagov2011,Barth2016}. We take standard GaAs parameters \cite{krummheuer2005pure} for a quantum dot with radius \SI{4}{nm} at a temperature of \SI{4}{K}. Choosing the 2C-SUPER scheme as in Fig.~\ref{fig:twocolor_example} and Tab.~\ref{tab:PIB}, we find that phonons hardly influence the scheme. To be specific, the final occupation without phonons is \SI{99.99}{\percent}, while the calculation with phonons yields an occupation  of \SI{97.31}{\percent} and thus the difference is negligibly small. Phonons might also lead to a degradation of the photon properties \cite{thoma2016exploring,Gustin2018,Cosacchi2021PRL}, which will be explored in future work. Nonetheless, the small effect of the phonons on the SUPER scheme for state preparation gives a good hint that our scheme can be used to generate high-quality photons.

\section{Conclusion}\label{sec:conclusion}
In this work we propose to use largely detuned pulses for the preparation of the excited state in a quantum emitter system using the SUPER scheme. The mechanism relies on the swing-up of the excited state occupation by combining Rabi oscillations with different frequencies, which alone would not lead to any significant occupation. The SUPER mechanism can be exploited either by using (i) a frequency modulation or (ii) a two-color scheme. 

In both cases, the excitation takes place in the transparent region of the material yielding a very distinct separation of exciting pulse and detection. The 2C-SUPER scheme can be experimentally realized using state-of-the-art laser pulses and results in the generation of high-quality single photons.

In the recent years, many different schemes for the state preparation of quantum emitters have been proposed and implemented that go beyond simple Rabi rotations \cite{luker2019review} like ARP \cite{simon2011robust,wei2014deterministic,kaldewey2017coherent} or phonon-assisted schemes \cite{Ardelt2014,bounouar2015phononassisted,quilter2015phononassisted,barth2016fast,cosacchi2019emission}. Only the phonon-assisted schemes are off-resonant, but they rely on an incoherent process, while our scheme uses a coherent excitation mechanism. Recently also two-color excitation schemes have been developed to excite the quantum dot \cite{he2019coherently,koong2021coherent}, underlining the need to remove the filtering to obtain a high photon yield. However, in these schemes, the energies of the two laser pulses were symmetric to the transition frequency, hence, one pulse was close to the region of higher excited states. In contrast, our proposed scheme uses two pulses which are both strongly negatively detuned and hence in the transparent region.

This makes the SUPER scheme highly attractive for applications in the field of quantum information technology. 
%%%%%%%%%%%%%%%%%%%%%%%%%%%  ACKNOWLEDGEMENT %%%%%%%%%%%%%%%%%%%%%%%%%%% 
\acknowledgements
We thank the German Research Foundation DFG for financial support through
the project 428026575. T.H. acknowledges financial support by the German Federal Ministry of Education and Research (BMBF) via the project ‘QuSecure’ (Grant No. 13N14876) within the funding program Photonic Research Germany.
%%%%%%%%%%%%%%%%%%%%%%%%%%%  APPENDIX %%%%%%%%%%%%%%%%%%%%%%%%%%% 
\appendix
\section{System Hamiltonian}
\noindent{}The system studied in this paper consists of a ground state $\ket{g}$, an excited state $\ket{x}$ separated by an energy $\hbar\omega_0$ and is driven by a time dependent term $\Omega(t)$ within the rotating wave approximation (RWA). The RWA is useful in this case, because the detunings considered are in the range of a few meV, while the energy difference of the two levels is in the eV range. For a quantum dot, the driving is given in the dipole moment approximation, where $\Omega$ is proportional to the product of the electric field $E(t)$ and the dipole moment with $\Omega(t)=dE(t)/\hbar$. The Hamiltonian describing this system reads
\begin{equation}
    H = \hbar\omega_0\ket{x}\bra{x} - \frac{\hbar}{2}\left(\Omega^*\ket{g}\bra{x} + \Omega\ket{x}\bra{g}\right).
\end{equation}
Due to the rotating wave approximation, the field is given with its complex time dependence as given in  Eqs.~\eqref{eq:electric_field} and \eqref{eq:two_color_laser}. From this, equations of motion can be obtained using the von-Neumann equation. Introducing the occupation of the excited state $f=\langle\ket{x}\bra{x}\rangle$ and the coherence (or polarization) $p = \langle\ket{g}\bra{x}\rangle$ leads to the Bloch equations
\begin{align}
    \begin{split}
        \frac{d}{dt}f &= \text{Im}(\Omega^*p),\\
        \frac{d}{dt}p &= -i\omega_0 p + \frac{i}{2}\Omega(1-2f)
    \end{split}
\end{align}
Standard numerical approaches for solving differential equations can then be used to obtain the temporal dynamics of the system. The use of a rotating reference frame is beneficial for the numerical integration, as it reduces the fast oscillations due to the electric field, so that a larger time-step can be chosen.\\
For illustration purposes we make use of the Bloch vector representation. For this, we change into a reference frame rotating with the laser frequency, resulting in a driving term which is real. In this case the equation of motion can be formulated using a cross product%the Bloch vector $\mathbf{r}$
\begin{equation}
    \frac{d}{dt}\mathbf{r} = \mathbf{\Omega}\cross \mathbf{r},\quad \mathbf{r}=\begin{pmatrix}\phantom{-}2\text{Re}(p)\\-2\text{Im}(p)\\2f-1\end{pmatrix},\,\mathbf{\Omega}=\begin{pmatrix}-\Omega_0\\0\\-\Delta\end{pmatrix},
\end{equation}
where $\mathbf{r}$ is the Bloch vector, rotating around a rotation axis $\mathbf{\Omega}$. $\Omega_0$ corresponds to the envelope of the field as given in Eq.~\eqref{eq:electric_field} and $\Delta$ is the detuning. Note that a clear picture using the Bloch vector is not possible for fields like the one given in Eq.~\eqref{eq:two_color_laser}, as due to the two frequencies, no rotating frame exists that results in a purely real driving term. Still, also for the two-color scheme the use of a rotating frame is beneficial for numerical purposes. %, as this effectively reduces the oscillation frequency of the electric field, allowing for larger integration timesteps

%%%%%%%%%%%%%%%%%%%%%%%%%%%%% APP B
\section{Phase dependence} \label{app:phase}
\begin{figure}
    \centering
    \includegraphics{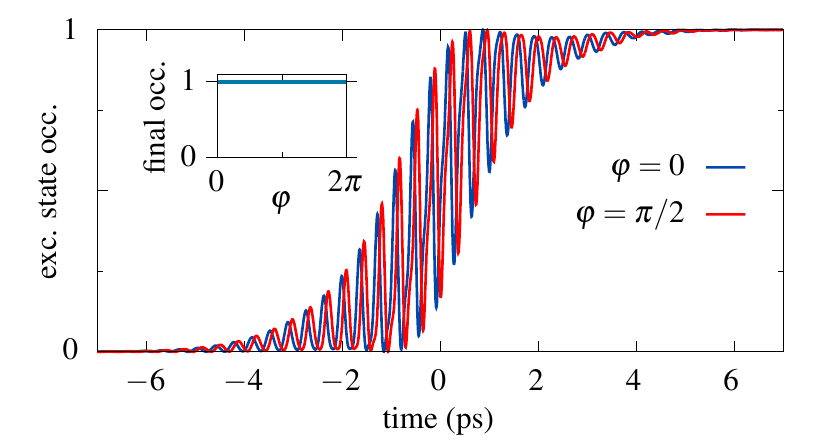}
    \caption{Comparison of the calculations for $\varphi=0$ in  Fig.~\ref{fig:twocolor_example} with the result for $\varphi=\pi/2$. The inset shows that the final occupation is constant for phases $\varphi=0,..,2\pi$.
    }
    \label{fig:twocolor_phase}
\end{figure}

In the 2C-SUPER scheme we introduced a relative phase $\varphi$ between the two pulses in  Eq.~\eqref{eq:two_color_laser}. When varying the phase from $\varphi=0,...,2\pi$, we find that the final occupation does not change as a function of $\varphi$. The results are given in Fig.~\ref{fig:twocolor_phase} for the same pulse parameters as in Fig.~\ref{fig:twocolor_example}. Exemplarily, we show the dynamics for $\varphi=0$ and  $\varphi=\pi/2$. Due to the phase, the oscillations are shifted slightly while the final occupation remains unchanged. This is true for all phases, the inset shows that the final excited state occupation for all possible relative phases is the same.

%%%%%%%%%%%%%%%%%%%%%%%%%%%%% APP C
\section{Behavior at very small detunings}
\label{app:small_det}
\begin{figure}
    \centering
    \includegraphics{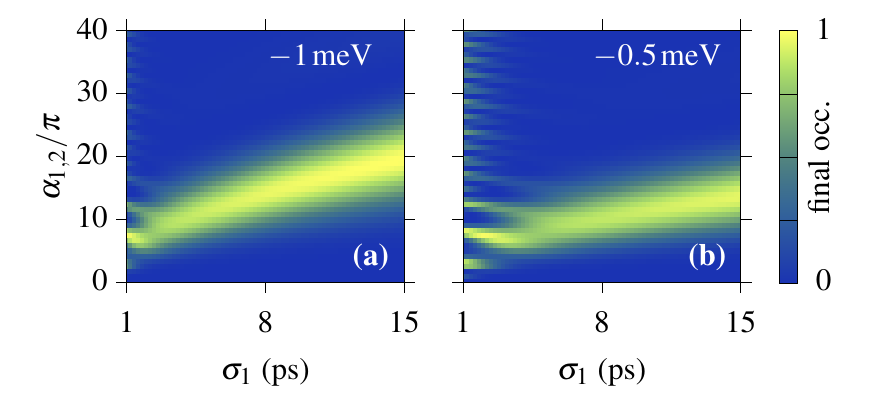}
    \caption{Final occupation of the excited state for the two-color scheme at lower detunings, with $\tau=0$ and $\sigma_2=1.5\sigma_1$. (a) $\hbar\Delta_1=\SI{-1}{meV}$, (b) $\hbar\Delta_1=\SI{-0.5}{meV}$
    }
    \label{fig:twocolor_small_det}
\end{figure}

The SUPER scheme works similarly well at smaller detunings ($\sim \SI{-1}{meV}$), however, at some point the spectrum of the pulse will overlap with the transition energy due to the spectral width of the pulses. This is shown in Fig.~\ref{fig:twocolor_small_det}, which is similar to Fig.~\ref{fig:multiparameter} but shows the results for smaller detunings of (a) $\hbar\Delta_1=\SI{-1}{meV}$ and (b) $\hbar\Delta_1=\SI{-0.5}{meV}$. In both cases, for small $\sigma_1$ a transition from 2C-SUPER to resonant Rabi-like oscillations occurs. For long pulses, i.e., when the spectral separation is larger, the occupation is achieved via the SUPER-scheme.
%%%%%%%%%%%%%%%%%%%%%%%%%%%%% APP D
\section{Definition of photonic quantities}
\label{app:def_photonic}
Here, we define the three photonic quantities discussed in Sec.~\ref{sec:single_photons}. The quantities are extracted from the projection operator $\hat{p}=\ket{g}\bra{x}$, yielding the polarization  with $p=\langle \hat{p} \rangle$ as used in the Bloch equations. \\
To obtain the correlation functions occurring in the following, a pulse train has to be modeled. The separation between the pulses in this train is denoted as $T_{\textrm{Pulse}}$ and chosen such that all quantities are relaxed before the next pulse hits the quantum dot. The brightness is estimated via the photon output $\O$, which we define as the number of photons emitted per excitation cycle given as \cite{Manson2016,Gustin2018,cosacchi2019emission}
\begin{align}
\label{eq:B}
\O:=\,&\gamma \int_{t_0-T_{\t{Pulse}}/2}^{t_0+T_{\t{Pulse}}/2} dt\, \<\hat{p}\+(t) \hat{p}(t)\>\, ,
\end{align}
where $t_0$ is the center time of the pulse and $0\leq\O\leq\g T_{\t{Pulse}}$.
We scale $\O$ such that $100\,\%$ corresponds to the ideal case of a delta-pulse excitation with pulse area $\pi$. Note that the photon output of the source is different to the number of photons collected per excitation, which might be significantly lower.

%\textbf{Single-photon purity}
The single-photon purity $\P$ is defined as
\begin{align}
\mathcal{P}=1-p_{\text{1ph}}\qquad\text{with} \qquad 
p_{\text{1ph}}=\frac{\int_{-T_{\t{Pulse}}/2}^{T_{\t{Pulse}}/2}d\tau\,G^{(2)}(\tau)}{\int_{T_{\t{Pulse}}/2}^{3 T_{\t{Pulse}}/2}d\tau\,G^{(2)}(\tau)}\, .
\end{align}
via the second-order correlation function 
%\begin{subequations}
\begin{align}
%\label{eq:G2_tau}
G^{(2)}(\tau):=\,&\lim_{T\to\infty}\frac{1}{T}\int_{0}^T dt\, G^{(2)}(t,\tau)\, , \nonumber\\
%\label{eq:G2_t_tau}
G^{(2)}(t,\tau):=\,&\<\hat{p}\+(t)\hat{p}\+(t+\tau)\hat{p}(t+\tau)\hat{p}(t)\> \nonumber
\end{align}
%\end{subequations}
\noindent{}The single-photon purity $\P$ is a measure for the single-photon component of the photonic state \cite{Michler2000,Santori2001,Santori2002,He2013,Somaschi2016,wei2014deterministic,Ding2016,Schweickert2018}. $\P=1$ implies a perfect single-photon purity. It has no lower bound, $-\infty<\P\leq1$, since $p_{\text{1ph}}$ can be larger than one in the case of bunching behavior instead of antibunching.

%\textbf{Indistinguishability}
The indistinguishability $\I$ of two successively emitted photons is
\begin{align}
\mathcal{I}=1-p_{\t{HOM}} \quad \text{with}\quad 
p_{\t{HOM}}=\frac{\int_{-T_{\t{Pulse}}/2}^{T_{\t{Pulse}}/2}d\tau\,G^{(2)}_{\t{HOM}}(\tau)}{\int_{T_{\t{Pulse}}/2}^{3 T_{\t{Pulse}}/2}d\tau\,G^{(2)}_{\t{HOM}}(\tau)}
\end{align}
with the correlation functions  \cite{Kiraz2004,Fischer2016,Gustin2018}
%\begin{subequations}
\begin{align}
%\label{eq:G2HOM_tau}
G^{(2)}_{\t{HOM}}(\tau):=\,&\lim_{T\to\infty}\frac{1}{T}\int_{0}^T dt\, G^{(2)}_{\t{HOM}}(t,\tau) \nonumber\\
%\label{eq:G2HOM_t_tau}
G^{(2)}_{\t{HOM}}(t,\tau):=\,&\frac{1}{2}\big[
\<\hat{p}\+(t) \hat{p}(t)\> \<\hat{p}\+(t+\tau) \hat{p}(t+\tau)\>\nn 
&-\big|\<\hat{p}\+(t+\tau)\hat{p}(t)\>\big|^2
+G^{(2)}(t,\tau)\big]\, . \nonumber
\end{align}
%\end{subequations}
The last term in $G^{(2)}_{\t{HOM}}(t,\tau)$ accounts for $\P\neq1$. Perfect indistinguishability corresponds to $\I=1$ and using the definition of $G^{(2)}_{\t{HOM}}(\tau)$ it is bounded by $0.5\leq\I\leq1$ \cite{Fischer2016}.

All correlation functions are calculated within the density matrix formalism by solving the Liouville-von Neumann equation and applying the quantum regression theorem for the propagation in the delay time argument \cite{Carmichael1993}.
Since the dynamics is purely Markovian in the case studied here, the quantum regression theorem is exact.

\bibliography{main}
\end{document}